\documentclass[preprint,showpacks,preprintnumbers,amsmath,amssymb]{revtex4}

\usepackage{graphicx}
\usepackage{dcolumn}
\usepackage{bm}

\begin{document}


\title{Nonlinear Software Sensor for Monitoring Genetic Regulation Processes with Noise and Modeling Errors}

\author{V. Ibarra-Junquera$^1$\footnote{E-mail: vrani@ipicyt.edu.mx, Fax: +52 444 834 2010}, L.A. Torres$^1$\footnote{E-mail: ltorres@ipicyt.edu.mx}, H. C. Rosu$^1$\footnote{E-mail: hcr@ipicyt.edu.mx, corresponding author}, G. Arg\"uello$^1$\footnote{E-mail: grarguel@ipicyt.edu.mx},\\ and \\ J. Collado-Vides$^2$\footnote{E-mail: collado@ccg.unam.mx} }

\affiliation{$^1$ Potosinian Institute of Science and Technology,\\
Apartado Postal 3-74 Tangamanga, 78231 San Luis Potos\'{\i}, Mexico\\
$^2$ Center of Genomic Sciences,
Universidad Nacional Autonoma de M\'exico, Apartado Postal 565-A Cuernavaca, Morelos 62100, Mexico}

\date{Received 18 Oct. 2004; revised manuscript received 3 Feb. 2005; published 29 July 2005}  %

\centerline{ArXiv: physics/0410096  [v3]}


\begin{abstract}
Nonlinear control techniques by means of a software sensor that are commonly used in chemical
engineering could be also applied to genetic regulation processes. We provide here a realistic formulation of this procedure
by introducing an additive white Gaussian noise, which is usually found in experimental
data. Besides, we include model errors, meaning that we assume we do not know the nonlinear regulation
function of the process.
In order to illustrate this procedure, we employ the Goodwin dynamics of the concentrations [B.C. Goodwin, \emph{Temporal Oscillations in Cells}, (Academic Press, New York, 1963)] in the simple form recently
applied to single gene systems and some operon cases [H. De Jong, J. Comp. Biol. {\bf 9}, 67 (2002)], which involves the dynamics of the mRNA, given protein, and metabolite concentrations.
Further, we present results for a three gene case in co-regulated sets of transcription units as they occur in prokaryotes.
However, instead of considering their full dynamics, we use only the data of the metabolites
and a designed software sensor. We also show, more generally, that it is possible to rebuild the complete set of nonmeasured concentrations despite
the uncertainties in the  regulation function or, even more, in the case of not knowing the mRNA dynamics. In addition, the rebuilding of concentrations is not affected by the perturbation due to the additive white Gaussian noise and also we managed to filter the noisy output of the biological system.\\  \\
\vspace*{10pt}
DOI: 10.1103/PhysRevE.72.011919 \hfill PACS
number(s): 87.10.+e, 05.45.-a
\end{abstract}

\vspace*{10pt}


\maketitle


\section{Introduction} 
Gene expression is 
a complex dynamic process with intricate regulation networks all along its stages leading to the
synthesis of proteins \cite{Lewin99}. Currently, the most studied aspect is that of regulation of initiation of transcription at the DNA level. Nevertheless,  the expression of
a gene product may be regulated at several levels, from transcription to RNA elongation and processing, RNA translation and even as post-translational modification of protein activity.
Control engineering is a key discipline with tremendous potential to simulate and manipulate the processes of gene expression.
In general, the control terminology and its mathematical methods are poorly known to the majority of biologists.
Many times the control ideas are simply reduced to the homeostasis concept. However,
the recent launching of the IEE journal {\em Systems Biology}  \cite{IEE} points to many promising
developments from the standpoint of systems analysis and control theory in biological sciences. Papers like that of Yi et al \cite{yi00}, in which the
Barkai and Leibler robustness model \cite{BL97} of perfect adaptation in bacterial chemotaxis is shown to have the property of a simple linear integral feedback control,
could be considered as pioneering work in the field.

We mention here two important issues. The first one is that the basic concept of state of a
system or process could have many different empirical meanings in biology. For the particular case of gene expression, the meaning of a state is essentially that of a
concentration. The typical problem in control engineering that appears to be tremendously useful in biology is the reconstruction of some specific regulated states under conditions of limited information. Moreover, equally interesting is the issue of noise filtering. It is quite well known that gene expression is a phenomenon with two sources of noise: one due to the inherent stochastic nature of the process itself and the other originating
in the perturbation of the natural signal due to the measuring device. In the mathematical approach, the latter class of noise is considered as an additive contamination of the real signal and this is also our choice here.
Both issues will form the subject of this investigation.

Taking into account the fact that rarely one can have a sensor on every
state variable, and some form of reconstruction from the available
measured output data is needed, a software can be constructed using
the mathematical model of the process to obtain an estimate
$\hat{X}$ of the true state $X$. This estimate can then be used as
a substitute for the unknown state $X$. Ever since the original work by
Luenberger \cite{Luenberger 1966}, the use of state observers
has proven useful in process monitoring and for many other tasks.
We will call herein as observer, in the sense of control theory, an algorithm capable of giving a reasonable estimation of
the unmeasured variables of a process.
For this reason, it is widely used in control, estimation, and
other engineering applications.

Since almost all observer designs are heavily based on
mathematical models, the main drawback is precisely the dependence
of the accuracy of such models to describe the naturally occurring processes.
Details such as model uncertainties and noise could affect the performance of the observers.
Taking into account these details is always an important matter and should be treated carefully.
Thus, we will pay special attention in this research to estimating unknown
states of the gene expression process under the worst possible case, which corresponds to noisy data, modeling errors, and unknown initial conditions.
These issues are of considerable interest and our approach is a novel contribution  to this important biological research area.
Various aspects of noisy gene regulation processes have been dealt with recently from both computational and experimental points of view in a number of interesting
papers \cite{noisyg}. We point out that since we add the noise $\delta$ to the output of the dynamic system in the form $y=CX+\delta$ (see Eqs.~$\Gamma$ in Section IV) it seems that
its origin is mainly extrinsic to the regulation process, even though it could be considered as a type of intrinsic noise with respect to the way the experiment is performed.
On the other hand, when writing the equation in the form $y=C(X +I\Delta)$, where $\Delta$ is a vector of noisy signals, one can see that the observer could estimate states that are intrinsically noisy even though
the processes are still deterministic.

\section{Brief on the biological context}  

Similar to many big cities, with heavy traffic, biological cells host complicated traffic of biochemical signals at all levels. Like cars on a busy highway, millions of molecules get involved in the bulk of the cell in many life processes controlled by genes. At the nanometer level, clusters of molecules in the form of proteins drive the dynamics of the cellular network that schematically can be divided into four regulated parts:  the DNA or
genes, the transcribed RNAs, the set of interacting proteins and the metabolites. Genes can only affect other genes through specific proteins, as well as through some metabolic pathways that are regulated by proteins themselves. They act to catalyze the information stored in DNA, all the way from the fundamental processes of transcription and translation to the final quantities of produced proteins.

Considering the enormous complexity of multicellular organisms generated by their large genomes, one can nevertheless still associate at least one regulatory element to any component gene. Each regulatory system is then composed of two elements at the DNA level, the gene that encodes a transcriptional regulator, and the target in the DNA where this regulator binds to, and excerts its activator or repressor function in transcription. These loops of interactions represent a fundamental piece to understand  the functioning of complex regulatory transcriptional and translational networks \cite{brief1, brief2}.
For the purpose of modelling, it is essential to generate simple models that help to understand elementary dynamical components of these complex regulatory networks as molecular tools that
participate in an important way in the machinery of cellular decisions, that is to say, in the behaviour and genetic program of cells.

Many entities in cellular networks can be identified as the basic units of regulation, mainly distinguished by their unique roles with respect to interaction with other units. These basic units are: the genes, with codifying content, also described as structural genes; the regulatory elements that in the old literature were called regulatory genes, which are smaller fragments of DNA sequences (of the order of 5 to 20 nucleotides) called operator sites where regulatory proteins as well as the RNA polymerase bind to; the messanger RNAs or mRNAs which are the products of transcription and form the template for the subsequent production of proteins as encoded by the corresponding gene; the forms of each protein and protein complexes, as well as, all metabolites present in the cell, either as products of enzymatic reactions or internalized by transport systems. These units have associated values that either represent concentrations or levels of activation. These values depend on both the values of the units that affect them due to the aforementioned mechanisms and on some parameters that govern each special form of interaction.

This gives rise to genetic regulatory systems structured by networks of regulatory interactions between DNA, RNA, proteins, and small molecules.
The simplest regulatory network is made of only one gene that is transribed into mRNA, this mRNA is then translated into proteins, which can be activated or inhibited as a result of their interaction with other proteins
or with specific metabolites. Transcriptional regulators are two-head structures, one being the domain of DNA interaction, and the other one is the so-called allosteric domain that interacts with specific metabolites.
Taking together these properties of the molecular machinery, one can envision that a gene encodes a protein which can regulat its own activity, either positively or negatively, depending on its effect in
enhancing or preventing the RNA polymerase transcriptional activity on its own gene by means of binding to an operator sites upstream of its own encoding gene. Upstream here meaning before the beginning of the gene where transcription initiates.
A mathematical model of such a biological inhibitory loop has been discussed since a long time ago by Goodwin and recurrently occurred in the literature, most recently being reformulated by De Jong \cite{DeJong02}. Although this case could look unrealistic, there are simple organisms, such as bacteria, where one regulatory loop may prove essential as recently discussed in detail by Ozbudak et al \cite{ozbudak}.
However, already at the level of two genes the situation gets really complicated, mostly because of the possible formation of
heterodimers between the repressors and other proteins around. These heterodimers are able to bind at the regulatory sites of the gene and therefore can affect it and lead to modifications of the regulatory process.

Recent development of experimental techniques, like cDNA microarrays and oligonucleotide chips, have allowed rapid measurements of the spatiotemporal expression levels of genes
\cite{briefbb99, lip99, LocWin00}. In addition, formal methods for the modeling and simulation of gene regulation processes are currently being developed in parallel to these experimental tools. As most genetic regulatory systems of interest involve many genes connected through interlocking positive and negative feedback loops, an intuitive understanding of their dynamics is hard to obtain. The advantage of the formal methods is that  the structure of regulatory systems can be described unambiguously, while predictions of their behavior can be made in a systematic way.

To make the description very concrete, it is interesting to look at well-defined, i.e., quite simple mathematical models that we present in the next section that refers to single gene cases and single gene clusters
(operons). The nonlinear software sensor for such cases is discussed in Section IV. A three-gene case is treated as an extension to regulatory gene networks and shows that the method of forward
engineering still works for reasonably simple gene networks. The conclusion section comes at the end of the paper.

\section{Mathematical Model for Gene Regulation}
In this section, we use the very first kinetic model of a genetic regulation process
developed by Goodwin in 1963 \cite{Goodwin63}, generalized by Tyson in 1978 \cite{Tyson78} and most recently
explained by De Jong \cite{DeJong02}. The
model  in its most general form is given by the following set of equations:
\begin{eqnarray}
\dot{X}_1 &=& {K}_{1n} r\left(X_{n} \right)-{\gamma}_{1}{X}_{1} \label{Eqn1}~,\\
\dot{X}_i &=& {K}_{i,i-1}X_{i-1}-{\gamma}_{i}{X}_{i}~, \
\quad 1<i\leq n~.    \label{Eqn3}
\end{eqnarray}
The parameters $K_{1n}, K_{21}, \ldots,K_{n,n-1}$ are all strictly
positive and represent production constants, whereas $\gamma
_1,\ldots, \gamma_n$ are strictly positive degradation
constants. These rate equations express a balance between the
number of molecules appearing and disappearing  per unit time. In
the case of $X_1$, the first term is the production term involving
a nonlinear nondissipative regulation function.
We take this as an unknown function. On the other hand, the
concentration $X_i$, $1<i \leq n$, increases linearly with
$X_{i-1}$. As well known, in order to express the fact that the metabolic
product is a co-repressor of the gene, the regulation function
should be a decreasing function for which most of the authors use
the Hill sigmoid, the Heaviside and the logoid curves. The
decrease of the concentrations through degradation, diffusion and
growth dilution is taken proportional to the concentrations
themselves. For further details of this regulation model we recommend  the reader the
review of De Jong \cite{DeJong02}.

It is to be mentioned here that
bacteria have a simple mechanism for coordinating the regulation of genes that encode products involved in a set of related processes: these genes are clustered on the chromosome and are transcribed together. Most prokaryotic mRNAs are polycistronic (multiple genes on a single transcript) and the single promoter that initiates transcription of clusters is the site of regulation for expression of all genes in the cluster. The gene cluster and promoter, plus additional sequences that function together in regulation, are called operon. Operons that include two to six genes transcribed as a unit are common in nature \cite{Lehninger}.

The fact that two or more genes are transcribed together on one polycistronic mRNA implies that we have a unique mRNA production constant and consequently we also have one mRNA degradation constant. In addition, the polycistronic mRNA can be translated into one or several enzymes, resulting in the existence of just one enzyme production and degradation constant, respectively. The same applies for the metabolite produced through the enzyme catalysis. Thus, if the resulting metabolite has repressor activity over the polycistronic mRNA (as in the case of tryptophan \cite{trypto}), then the model given by Eqs.~(\ref{Eqn1},\ref{Eqn3}) could also be applied to operons and therefore it has a plausible application to the study of prokaryotic gene regulation.

\section{Nonlinear Software Sensor}

Numerous attempts have been made to develop nonlinear observer
design methods. One could mention the industrially popular
extended Kalman filter, whose design is based on a local
linearization of the system around a reference trajectory,
restricting the validity of the approach to a small region in
the state space \cite{Stephanopoulos,Wolovich}. The first
systematic approach for the development of a theory of nonlinear
observers was proposed some time ago by Krener and Isidori
\cite{Krener 83}. In further research, nonlinear transformations of
the coordinates  have also been employed to put the considered
nonlinear system in a suitable ``observer canonical form'', in
which the observer design problem may be easily solved \cite{J.P.
Gauthier 92,J.P. Gauthier 91,J.P. Gauthier 94}.
Nevertheless, it is well known that classical proportional
observers tend to amplify the noise of on-line measurements, which
can lead to the degradation of the observer performance. In order to
avoid this drawback, this observer algorithm is based on the
works of Aguilar et al. \cite{R. Aguilar y col. 2003,R.
Aguilar 2003}, because the proposed integral observer provides
robustness against noisy measurement and uncertainties. We show
that this new structure retains all the characteristics of the
popular (the traditional high gain) state observers of the
classical literature and furthermore provides additional
robustness and noise filtering and thus can result in a
significant improvement of the monitoring performances of the
genetic regulation process.

\noindent
In this section, we present the design of a nonlinear
software sensor in which one $X_{j}$, for $j \in (1,...,n)$, is
the naturally measured state (the most easy to measure).
Therefore, it seems logical to take $X_{j}$ as the output of the
system
\begin{eqnarray}
y=h(X)=X_{j}~.
\end{eqnarray}
Now, considering the constant ${K}_{1n}$ and the function
$r\left(X_{n} \right)$ as unknown, we group them together in a function
$\Im(X)$. In addition, we consider that the output function $h(X)$ is
contaminated with a Gaussian noise. In such a case, the model given by the
aforementioned Eqs.~(\ref{Eqn1}) and (\ref{Eqn3}), acquires the
form:
\begin{eqnarray}
\Gamma: \ \ \ \left \{
\begin{array}{c}
\dot{X} = \bar{\Im}(X) + \ell(X)\\
 y = C X +\delta\\
\end{array}
\right. \nonumber
\end{eqnarray}
\noindent
where $\bar{\Im}(X)$ is a $n \times 1$ vector whose first
entry is ${\Im}(X)$ and all the rest are zero, $\ell(X)$ is also a $n
\times 1$ vector of the form $[-{\gamma}_{1}{X}_{1},
{K}_{i,i-1}X_{i-1}-{\gamma}_{i}{X}_{i}]^T$, $\delta$ is an additive
bounded measurement noise, and  $X\in \mathbb{R}^{n}$. The system is assumed to lie in
a ``physical subset'' $\Sigma \subset
\mathbb{R}^{n}$.

Then, the task of designing an observer for the system $\Gamma$ is to
estimate the vector of states $X$, despite of the unknown
part of the nonlinear vector $\bar{\Im}(X)$ (which should be
also estimated) and considering that $y$ is measured on-line and
that the system is observable.

\begin{figure}[h]
\centering
\includegraphics[scale=0.25]{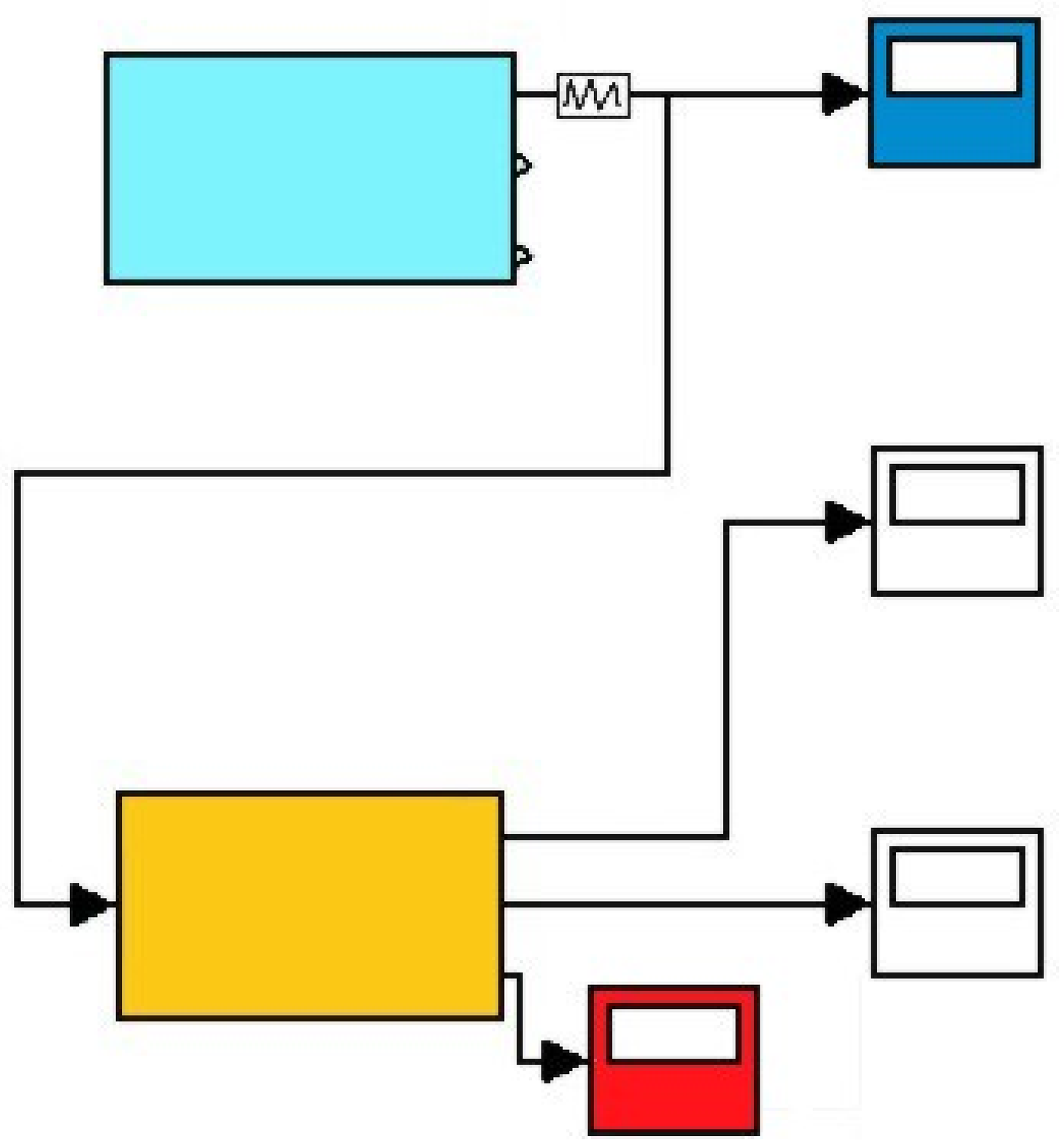}
    \put(-40,108){\tiny{Metabolite K}}
     \put(-37,101){\tiny{(unfiltered)}}
    \put(-34,57){\tiny{Protein A}}
    \put(-29,13){\tiny{mRNA}}
    \put(-82,122){\tiny{${X}_3$}}
    \put(-77, 135){\tiny{ Additive}}
    \put(-69, 127){\tiny{Noise}}
    \put(-84,34){\tiny{$\hat{X}_2$}}
    \put(-84,25){\tiny{$\hat{X}_1$}}
    \put(-135,90){\small{Original Process}}
    \put(-115,5){\small{Software}}
    \put(-110,-5){\small{Sensor}}
    \put(-72,-7){\tiny{Metabolite K}}
    \put(-66,-12){\tiny{(filtered)}}
\caption{Schematic representation of the software sensor, where
the output of the system is the input of the software sensor and
the outputs of the latter are the rebuilt concentrations.}
\label{Fig2d}
\end{figure}

A particular representation of the software sensor that we describe
here is provided in Fig.~\ref{Fig2d}.

In order to provide the observer with robust properties against
disturbances, Aguilar and collaborators \cite{R. Aguilar y col. 2003} considered only an integral type
contribution of the measured error. Moreover, an uncertainty estimator
is introduced in the  methodology of observation with the purpose of estimating
the unknown components of the nonlinear vector $\bar{\Im}(X)$. As a result,
the following representation of the system is proposed

\begin{eqnarray}
\Xi: \ \ \ \left \{
\begin{array}{c}
\dot{X}_{0}= C X +\delta\\
\dot{X} = \bar{\Im} + \ell(X)\\
\dot{\bar{\Im}}=\Theta (X)\\
 y_{0} = X_{0}
\end{array}
\right. \nonumber
\end{eqnarray}

that is, in the case of the model given by Eqs.~(\ref{Eqn1}) and (\ref{Eqn3})
\begin{eqnarray}
\dot{X}_0 &=& X_j+\delta \nonumber\\
\dot{X}_1 &=& X_{n+1}-{\gamma}_{1}{X}_{1} \nonumber\\
\dot{X}_i &=& {K}_{i,i}X_{i-1}-{\gamma}_{i}{X}_{i}~, \quad 1<i\leq n~,\nonumber\\
\dot{X}_{n+1} &=& \Omega(X)\nonumber\\
y &=& X_0~,
\end{eqnarray}
where $\dot{X_0}$ is the dynamical extension that allows us to integrate the noisy signal in order to recover a filtered signal, while $\dot{X}_{n+1}$ allows us to put the unknown  regulation function as a new state.
Thus, the task becomes the estimation of this new state (a standard task for an observer), and therefore the function $\Omega$ is related to the unknown dynamics of the new state.
At this point, $X \in \mathbb{R}^{n+2}$, and furthermore the following equation is
generated
$$
\dot{X}=AX+B+E\delta ~,
$$
where $AX$ is the linear part of the previous system such that $A$ is a matrix equivalent in form to a Brunovsky matrix,  
$B= [0, \ldots ,0,\Omega(X)]^T$ and $E=[1,0,\ldots,0]^T$.

We will need now the following result proven in Ref.~\cite{R.
Aguilar y col. 2003}.

{\small \em An asymptotic-type observer of the system $\Xi$ is given as follows:
\begin{eqnarray}
\hat{\Xi}: \ \ \ \left \{
\begin{array}{c}
\dot{\hat{X}}_{0}= C \hat{X} +\theta_1 \left(y_{0}- \hat{y}_{0}\right)\\
\dot{\hat{X}} = \hat{\bar{\Im}} + \ell(\hat{X})+ \theta_2 \left(y_{0}-\hat{y}_{0}\right)\\
\dot{\hat{\bar{\Im}}}=\theta_3 \left(y_{0}-\hat{y}_{0}\right)\\
 \hat{y}_{0} = \hat{X}_{0}~,
\end{array}\right. \nonumber
\end{eqnarray}}

where the gain vector $\theta$ of the observer is given by
\begin{eqnarray}
\theta= S_{\theta}^{-1} C^T~, \nonumber\\
S_{\theta;i,j}=\left(\frac{S_{i,j}}{\vartheta^{i+j+1}}\right)~.
\nonumber
\end{eqnarray}
Each entry of the matrix $S_\theta$ is given by the above equation, where $S_{\theta}$ is a $n \times n$ matrix ($i$ and $j$ run from 1 to $n$), and $S_{i,j}$ are entries of a symmetric positive definite matrix that do not depend on $\vartheta$. Thus, $S_{i,j}$  are such that $S_{\theta}$ is a positive solution of the algebraic Riccati equation
\begin{eqnarray}
S_{\theta}
\left(A+\frac{\vartheta}{2}I\right)+\left(A+\frac{\vartheta}{2}I\right)
S_{\theta} = C^T C~. \label{Ricatti}
\end{eqnarray}
In all formulas, $C = [1,0, ...,0]$. In the multivariable case we must create one matrix $S_{\theta}$ for each block corresponding to each output.
It is worth mentioning that we can think about this observer as a `slave' system that follows the `master' system, which is precisely the real experimental system. In addition,
$S_{\theta}$, as functional components of the gain vector, guarantees the accurate estimation of the observer through the convergence to zero of the error dynamics, i.e., the dynamics of the difference between the measured state and its corresponding estimated state. 
One can see that $\vartheta$ generates an extra degree of freedom that can be tuned by the user such that the performance of the software sensor becomes satisfactory for him.

In \cite{Martínez-Guerra 1994} it has been shown that such an
observer has an exponential-type decay for any initial conditions.
Notice that a dynamic extension is generated by considering the
measured output of the original system as new additional dynamics
with the aim to filter the noise. This procedure eliminates most of
the noise in the new output of the system. The reason of the
filtering effect is that the dynamic extension acts at the level of
the observer as an integration of the output of the original system,
(see the first equation of the system $\Xi$ and the error part in
the equations of system $\hat{\Xi}$). The integration has averaging
effects upon the noisy measured states. More exactly, the difference
between the integral of the output of the slave part of system
$\hat{\Xi}$ and the integral of the output of the original system
gives the error and the observer is planned in such a way that the
error dynamics goes asymptotically to zero, which results in the
recovering of both the filtered state and the unmeasured states.

\subsection{Particular Case}

\begin{figure}[h]
\centering
\includegraphics[height=7cm]{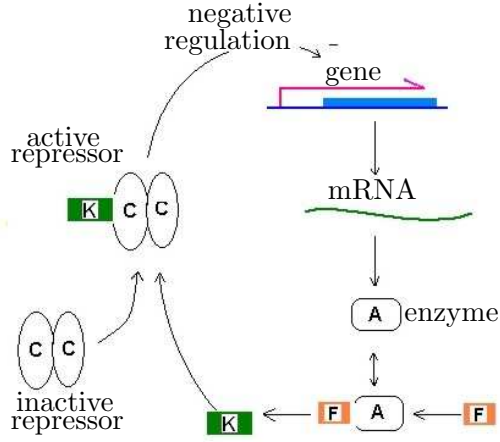}
    \put(-192,192){negative}
    \put(-201,182){regulation}
    \put(-257,45){inactive}
    \put(-259,38){repressor}
    \put(-253,145){active}
    \put(-259,139){repressor}
    \put(-140,170){gene}
    \put(-139,125){mRNA}
    \put(-110,79){enzyme}
\caption{The genetic regulatory system given by Eqs.~(\ref{Eqn1obs}) - (\ref{Eqn3obs}) involving end-product
inhibition according to De Jong \cite{DeJong02}. A is an enzyme
and C a repressor protein, while K and F are metabolites. The
mathematical model, as used by De Jong and by us, takes into
account experiments where only metabolite K is measured.} \label{Fig2}
\end{figure}

For gene regulation processes, which are of interest to
us here, we merely apply the aforewritten system of equations corresponding to the asymptotic observer $\hat{\Xi}$
\begin{eqnarray}
\dot{X}_1 &=& {K}_{1,3}r\left(X_3\right)-{\gamma}_{1}{X}_{1} \label{Eqn1obs}\\
\dot{X}_2 &=& {K}_{2,1}X_1-{\gamma}_{2}{X}_{2}\label{Eqn2obs}\\
\dot{X}_3 &=& {K}_{3,2}X_2-{\gamma}_{3}{X}_{3}~.\label{Eqn3obs}
\end{eqnarray}
The pictorial representation of this system of equations is given in
Fig.~\ref{Fig2}.

The values of the parameters given in Table~1, without
necessarily being the experimental values, are however consistent with the
requirements of the model.

\begin{table}[htb]
\centering \caption{Parameters of the model} \label{tab:1}
\begin{tabular}{lll}
\hline\noalign{\smallskip}
Symbol & \ \ \ \ \ \ \ \ \ \ \ Meaning & \ \ \ \ Value   \\
            & \ \ \ \ \ \ \ \ \ \             & (arb. units) \\
\noalign{\smallskip}\hline\noalign{\smallskip}

$K_{1,3}$ & Production constant of mRNA &\ \ \ \ \ $0.001$   \\
$K_{2,1}$ & Production constant of protein A &\ \ \ \ \ $1.0$  \\
$K_{3,2}$ & Production constant of metabolite K &\ \ \ \ \  $1.0$  \\
$\gamma_1$ & Degradation constant of mRNA &\ \ \ \ \ $0.1$  \\
$\gamma_2$ & Degradation constant of protein A &\ \ \ \ \ $1.0$  \\
$\gamma_3$ & Degradation constant of metabolite K &\  \ \ \ \ $1.0$  \\
$\vartheta$ & Hill's threshold parameter &\ \ \ \ \ $1.0$  \\
\noalign{\smallskip}\hline
\end{tabular}
\end{table}

Using the structure given by the equations of $\hat{\Xi}$, the
explicit form of the software sensor is:
\begin{eqnarray}
\dot{\hat{X}}_0 &=& \hat{X}_3+\theta_1 (y_0-\hat{X}_3) \nonumber\\
\dot{\hat{X}}_1 &=& X_{4}-{\gamma}_{1}{X}_{1}+\theta_2 (y_0-\hat{y}_0)\nonumber\\
\dot{\hat{X}}_2 &=& {K}_{2,1} X_1 - \gamma_{2}{X}_{2}+\theta_3 (y_0-\hat{y}_0)\nonumber\\
\dot{\hat{X}}_3 &=& {K}_{3,2}X_2-{\gamma}_{3}{X}_{3}+\theta_4 (y_0-\hat{y}_0)\nonumber\\
\dot{\hat{X}}_4 &=& \theta_5 (y_0-\hat{X}_3)~,\nonumber\\
\hat{y}_0 &=& \hat{X_0}~.\nonumber
\end{eqnarray}
Notice that this dynamic structure does not involve the regulation function.

\begin{figure}
\centering
      \includegraphics[height=5.3cm]{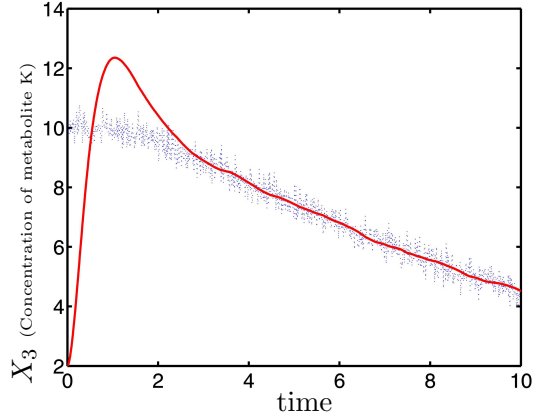}
      \put(-98,-7){time}
      \put(-198,3){\rotatebox{90}{$X_3$ \tiny{(Concentration of metabolite K)}}}
  \caption{Numerical simulation: solid lines represent the filtered states and the dotted
  lines represent the noisy measured state for the evolution in time of metabolite K concentration. Notice that the initial bad estimation
is due to the initial conditions that have been chosen far away from the real ones. This behaviour could be improved with a better knowledge of the initial conditions.
The units of the two axes are arbitrary, i.e., the model is nondimensional.}\label{filt}
\end{figure}

We can solve Eq.~(\ref{Ricatti}) and for numerical purposes we
choose $\vartheta=2.5$ and the standard deviation of the Gaussian
noise of 0.001. Figure~\ref{filt} shows the numerical simulation
that illustrates the filtering effect of the software sensor over
the noisy measured state.

\begin{figure}[htb]
  \hfill
  \begin{minipage}[t]{.45\textwidth}
    \begin{center}
      \includegraphics[height=4.65cm]{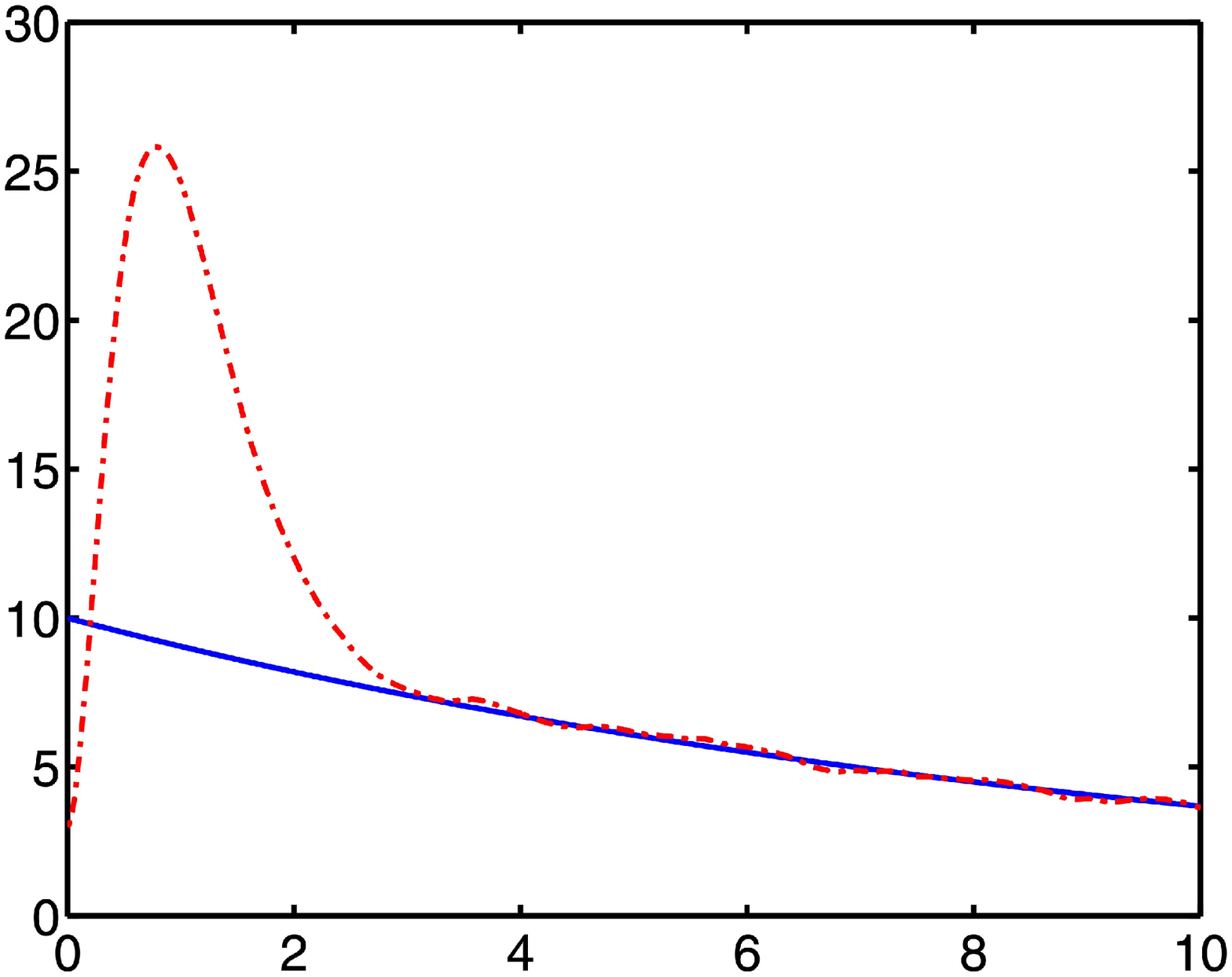}
      \put(-90,-10){time}
       \put(-30,110){(a)}
      \put(-175,10){\rotatebox{90}{$X_1$ \tiny{(concentration of mRNA)}}}
    \end{center}
  \end{minipage}
  \hfill
  \begin{minipage}[t]{.45\textwidth}
    \begin{center}
      \includegraphics[height=4.60cm]{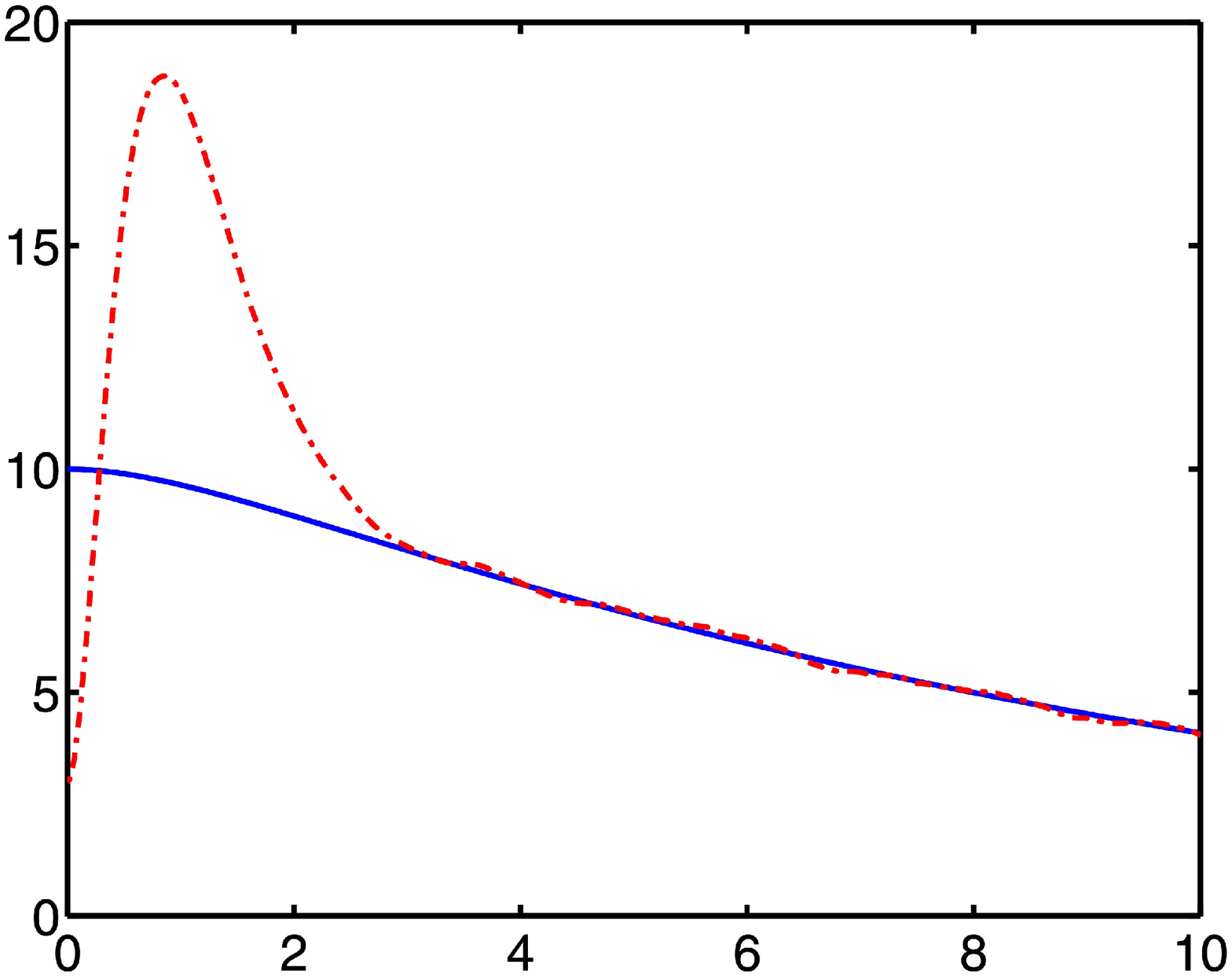}
       \put(-90,-10){time}
       \put(-170,10){\rotatebox{90}{$X_2$ \tiny{(concentration of protein A)}}}
       \put(-30,110){(b)}
    \end{center}
  \end{minipage}
  \hfill
  \caption{Numerical simulation: solid lines represent the true states generated by the original
process endowed with the Hill regulatory function and dotted lines represent the estimated concentrations provided by the software sensor
without any knowledge about the regulatory function. Plot (a) represents the evolution of mRNA
   concentration in time and plot (b) the variation of the concentration of
   protein A in time. The two axes have arbitrary units.}\label{simul}
\end{figure}

On the other hand,  Fig.~\ref{simul} shows the results of a
numerical simulation, where the solid lines stand for the true
states and the dotted lines indicate the estimates, respectively.

\newpage

\section{Three-Gene Circuit Case}

\begin{figure}[h]
\centering
\includegraphics[scale=0.82]{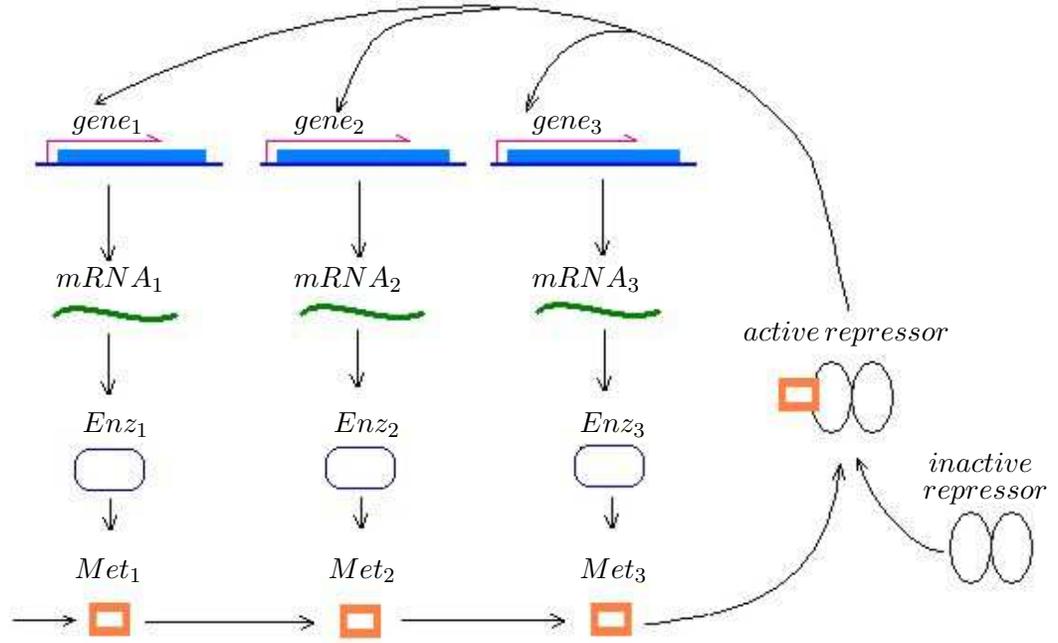}
\put(-374,225){\small{$gene_1$}}
\put(-290,225){\small{$gene_2$}}
\put(-200,225){\small{$gene_3$}}
\put(-380,165){\small{$mRNA_1$}}
\put(-290,165){\small{$mRNA_2$}}
\put(-200,165){\small{$mRNA_3$}}
\put(-120,145){\small{$active\, repressor$}}
\put(-50,95){\small{$inactive$}}
\put(-52,85){\small{$repressor$}}
\put(-370,110){\small{$Enz_1$}}
\put(-275,110){\small{$Enz_2$}}
\put(-182,110){\small{$Enz_3$}}
\put(-373,55){\small{$Met_1$}}
\put(-277,55){\small{$Met_2$}}
\put(-182,55){\small{$Met_3$}}
\caption{The three-gene regulatory circuit under consideration.}
\label{figureXX}
\end{figure}

In this section we extend the previous results to a more complicated case that can occur in prokaryotic cells. We study a more elaborated system where one regulator affects different promoters and transcription units. The case corresponds to the coupled regulation of three genes in which the metabolite resulting from the translation of gene 1 becomes the substrate for the synthesis of the metabolite catalyzed by the enzyme translated from gene 2, and similarly for gene 3, but the metabolite 3 becomes the repressor of all the three genes involved, as shown in Fig.~(\ref{figureXX}).

In this case the model is given by an extension of the model given by Eqs.~(\ref{Eqn1},\ref{Eqn3}). That results in the following system of differential equations:
\begin{eqnarray}
\frac{d}{dt}[mRNA_1] &=& K_1 R([Met_3]) -\gamma _1 [mRNA_1] \nonumber \\
\frac{d}{dt} [Enz_{1}] &=& K_2 [mRNA_1]-\gamma _2 [Enz_1] \nonumber \\
\frac{d}{dt}[Met_{1}] &=& K_3 [Enz_1] -\gamma _3 [Met_1]- \alpha _1 [Enz_2] \nonumber\\
\frac{d}{dt}[mRNA_{2}]&=& K_4 R(Met_3) -\gamma _4 [mRNA_2] \nonumber\\
\frac{d}{dt}[Enz_{2}]&=& K_5 [mRNA_2]-\gamma _5 [Enz_2] \nonumber\\
\frac{d}{dt}[Met_{2}]&=& K_6 [Enz_2] -\gamma _6 [Met_2] - \alpha _2 [Enz_3] \nonumber\\
\frac{d}{dt}[mRNA_{3}]&=& K_7 R([Met_3]) -\gamma _7 [mRNA_3] \nonumber\\
\frac{d}{dt}[Enz_{3}]&=& K_8 [mRNA_3]-\gamma _8 [Enz_3] \nonumber\\
\frac{d}{dt}[Met_{3}]&=& K_9 [Enz_3] -\gamma _9 [Met_3] ~,  \nonumber
\end{eqnarray}
where $[mRNA_i]$, $[Enz_i]$ and $[Met_i]$ represent the concentration of mRNA, enzymes and metabolites for each gene respectively. We select as the measured variables the metabolites because we want to show that through the measurement of stable molecules such as the metabolites, it is possible to infer the concentration of unstable molecules such as the mRNAs. Note that the equations are coupled through the dynamics of the metabolites. Moreover, we will assume that the dynamics of mRNA is bounded but unknown.

As we showed in the previous sections our new system can be written as:

\begin{eqnarray}
\dot{X}_1   &=&  X_2  + d_1\\
\dot{X}_2   &=&  K_3 X_3 - \gamma_3 X_2 - \alpha_1 X_8\\
\dot{X}_3   &=&  K_2 X_4 - \gamma_2 X_3\\
\dot{X}_4   &=&  X_5\\
\dot{X}_5   &=&  \phi _1 (X)\\
\dot{X}_6   &=&  X_7  + d_2\\
\dot{X}_7   &=&  K_6 X_8 - \gamma_6 X_7- \alpha_2 X_{13}\\
\dot{X}_8   &=&  K_5 X_9 - \gamma_5 X_8\\
\dot{X}_9   &=&  X_{10}\\
\dot{X}_{10} &=&  \phi _2 (X)\\
\dot{X}_{11} &=& X_{12}  + d_3\\
\dot{X}_{12} &=& K_9 X_{13} - \gamma_9 X_{12}\\
\dot{X}_{13} &=& K_8 X_{14} - \gamma_8 X_{13}\\
\dot{X}_{14} &=& X_{15}\\
\dot{X}_{15} &=& \phi _3(X)~,
\end{eqnarray}
where $mRNA_1= \dot{X}_{4}$,   $mRNA_2= \dot{X}_{9}$,   $mRNA_3= \dot{X}_{14}$ , $Enz_1= \dot{X}_{3}$,   $ Enz_2= \dot{X}_{8}$,   $ Enz_3= \dot{X}_{13}$,    $Met_1= \dot{X}_{2}$,   $ Met_2= \dot{X}_{7}$,
$Met _3= \dot{X}_{12}$,  $d_i$ represent the noise, $\phi _i(X)$ stand for the unknown dynamics. In adition, the previous systems can be written in the matricial forma as:
\begin{eqnarray}
\dot{X}   &=&  \overline{A} X  + \overline{B} (X) + E d , \ X\in \mathbb{R}^n \nonumber\\
    y &=& \overline{C} X= \left( C_1 X^1 \  \ldots \ C_m X^m  \right)^T~,
\end{eqnarray}
where in this case  $X^i \in \mathbb{R}^{\lambda_i}$ is the $i$th
partition of the state $X$ so that  $X=  [ ( X^1 )^T, \ldots, (
X^m )^T ]^T$ and $\sum_{i=1}^{m} {\lambda}_{i} = n $;
$\overline{A} = {\rm diag} [A^1, \ldots, A^m ]$ where $A^i$ is
$\lambda_i \times \lambda_i$ such that $S_{\theta}^i$ in the
equation (\ref{Ricatti}) is invertible; $C=$
${\rm diag}  [ C_1, \ldots, C_m  ] $, where $C_i = [1, 0,\ldots
,0]$ $\in \mathbb{R}^{\lambda_i}$; $B(X)= {\rm diag} [B^1(X
)^T, \ldots, B^m (X )^T ]^T$;  $ E = {\rm diag} [E_1, \ldots,
E_m ] $, where $ E_i = [1, 0,\ldots ,0]$ $\in
\mathbb{R}^{\lambda_i}$.


According to the scheme presented in the previous section we construct an observer through the following system of differential equations

 \begin{eqnarray}
\dot{\hat{X}}_1   &=&  \hat{X}_2  + \theta _{11}(X_1-\hat{X}_1)\\
\dot{\hat{X}}_2   &=&  K_3 \hat{X}_3 - \gamma _3 \hat{X}_2 - \alpha _1 \hat{X}_8+ \theta _{12}(X_1-\hat{X}_1)\\
\dot{\hat{X}}_3   &=&  K_2 \hat{X}_4 - \gamma _2 \hat{X}_3+ \theta _{13}(X_1-\hat{X}_1)\\
\dot{\hat{X}}_4   &=&  \hat{X}_5+ \theta _{14}(X_1-\hat{X}_1)\\
\dot{\hat{X}}_5   &=&   \theta _{15}(X_1-\hat{X}_1)\\
\dot{\hat{X}}_6   &=&  \hat{X}_7  + \theta _{21}(X_6-\hat{X}_6)\\
\dot{\hat{X}}_7   &=&  K_6 \hat{X}_8 - \gamma _6 \hat{X}_7- \alpha_2 \hat{X}_{13}+ \theta _{22}(X_6-\hat{X}_6)\\
\dot{\hat{X}}_8   &=&  K_5 \hat{X}_9 - \gamma _5 \hat{X}_8+ \theta_ {23}(X_6-\hat{X}_6)\\
\dot{\hat{X}}_9   &=&  \hat{X}_{10}+ \theta _{24}(X_6-\hat{X}_6)\\
\dot{\hat{X}}_{10} &=& + \theta _{25}(X_6-\hat{X}_6)\\
\dot{\hat{X}}_{11} &=&  \hat{X}_{12}  + \theta \gamma _9 \hat{X}_{12}+ \theta _{32}(X_{11}-\hat{X}_1)\\
\dot{\hat{X}}_{13} &=& K_8 \hat{X}_{14} - \gamma_8 \hat{X}_{13}+ \theta _{33}(X_{11}-\hat{X}_{11})\\
\dot{\hat{X}}_{14} &=& \hat{X}_{15}+ \theta _{34}(X_1-\hat{X}_1)\\
\dot{\hat{X}}_{15} &=&  \theta _{35}(X_{11}-\hat{X}_{11})~,
\end{eqnarray}
where $\theta_i$ stand for the observer gain values. Note, that
this extension is not a direct application of that developed by
Aguilar et al. \cite{R. Aguilar y col. 2003}
in the sense that this is a extension to the multivariable case. In
addition, the matrix ${A}_{i} $ is equivalent to a matrix of
Brunovsky form, which guarantees the existence, uniqueness and
invertibility of the matrix solution ${S_{\theta}}^i$ \cite{Hermann
77}.  (The existence and the uniqueness of ${S_{\theta}}^i$ follows
from the facts that $-\frac{{\theta}_{i}}{2}  I - A_i$ is of
Hurwitz-type and that the pair $\left (-\frac{{\theta}_{i}}{2}I -
A_i, C_i\right)$ is observable \cite{Shim}).

\begin{figure}[h]
\centering
      \includegraphics[height=5.3cm]{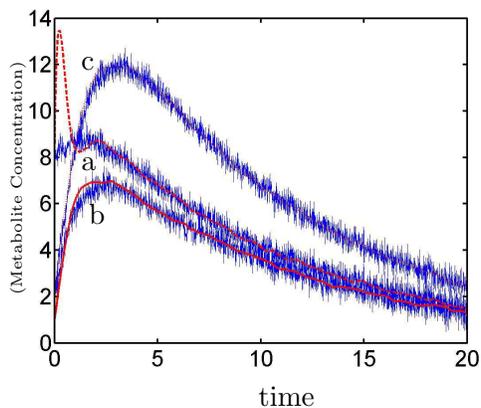}
      \put(-98,-7){time}
      \put(-165,120){c}
      \put(-165,82){a}
      \put(-162,62){b}
      \put(-192,35){\rotatebox{90}{\tiny{(Metabolite Concentration)}}}
  \caption{Numerical simulation: solid lines represent the filtered states
  obtained from the noisy measured states for the evolution in time of metabolite concentrations, where a, b and c correspond to metabolite 1, 2, and 3, respectively.
The units of the two axes are arbitrary (nondimensional model).}\label{ruido3}
\end{figure}

\begin{figure}[h]
  \hfill
  \begin{minipage}[t]{.45\textwidth}
    \begin{center}
      \includegraphics[height=4.65cm]{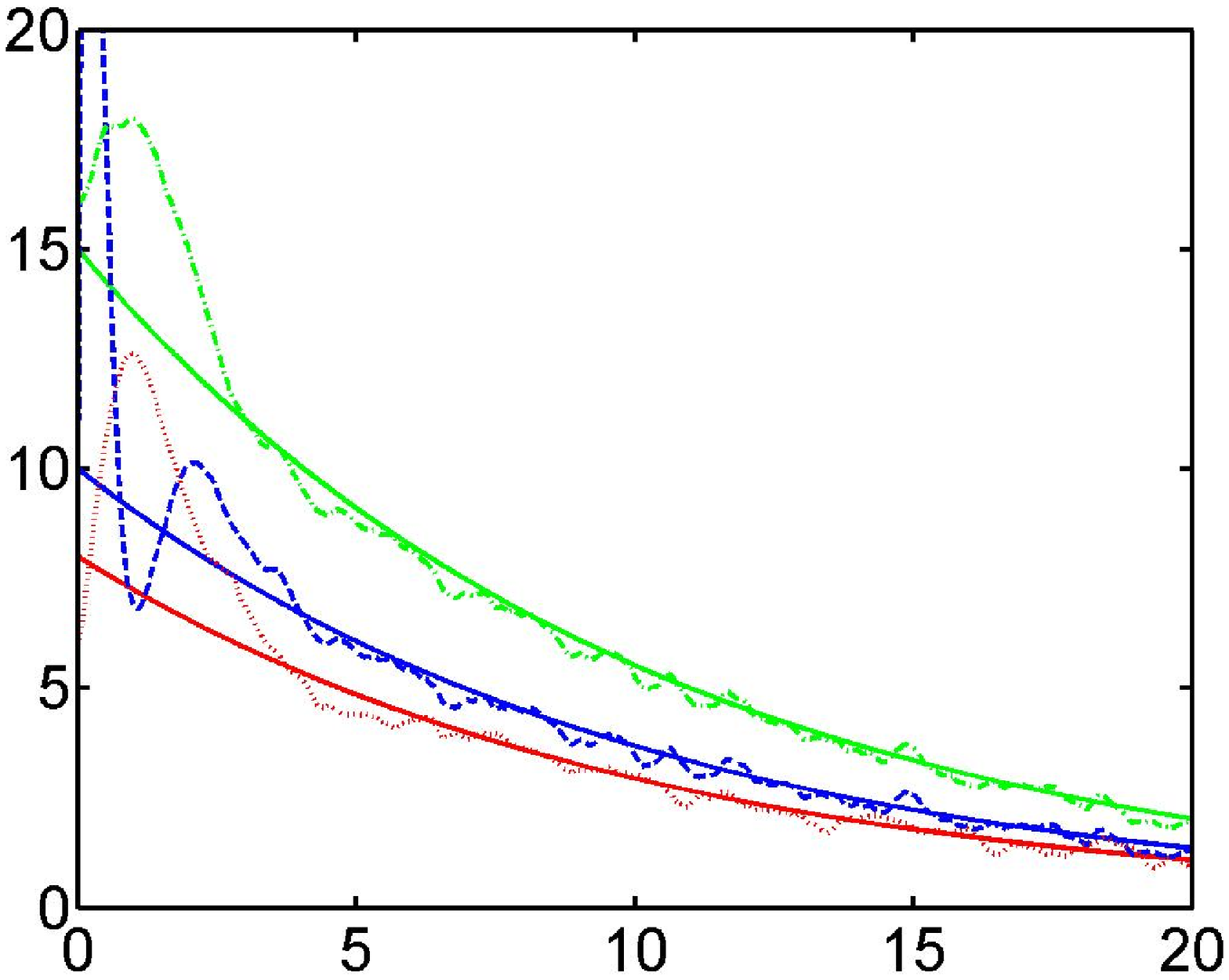}
      \put(-90,-10){time}
       \put(-40,110){($a$)}
       \put(-120,65){c}
       \put(-120,49){a}
       \put(-120,30){b}
      \put(-175,10){\rotatebox{90}{\tiny{(concentration of mRNA)}}}
    \end{center}
  \end{minipage}
  \hfill
  \begin{minipage}[t]{.45\textwidth}
    \begin{center}
      \includegraphics[height=4.60cm]{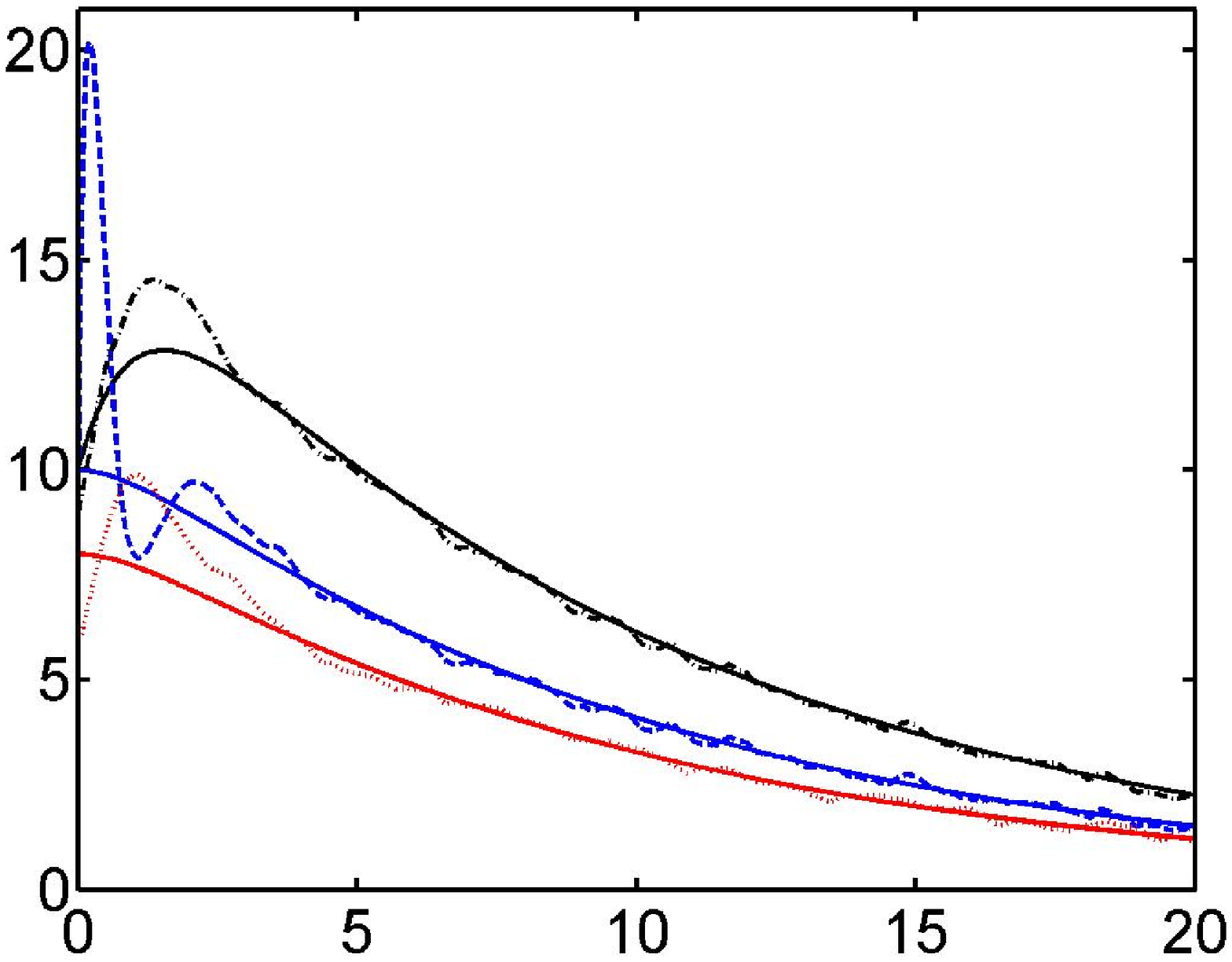}
       \put(-90,-10){time}
       \put(-170,10){\rotatebox{90}{\tiny{(concentration of enzymes)}}}
       \put(-40,110){($b$)}
       \put(-120,70){c}
       \put(-120,53){a}
       \put(-120,30){b}
    \end{center}
  \end{minipage}
  \hfill
  \caption{Numerical simulation: solid lines represent the true states generated by the original
process endowed with the Hill regulatory function and dotted lines
represent the estimated concentrations provided by the software
sensor without any knowledge about the regulatory function;
a, b and c correspond to molecule 1, 2 and 3, respectively.
Plot ($a$) represents the evolution of $mRNA_i$ concentrations in time and plot ($b$) the variation of the concentration of
the corresponding enzymes in time.The axes of the graph have arbitrary units.}\label{simul1}
\end{figure}

Figure~\ref{ruido3} shows the numerical simulation of the filtering
effect of the software sensor over the noisy measured state in this
case. On the other hand,  Fig.~\ref{simul1} displays the results of
a numerical simulation of the true states (solid lines) and the
estimates (dotted lines).



\section{Conclusion}

In this research, a simple software sensor was designed for a schematic gene regulation dynamic process involving end-product inhibition in single gene, operon and three gene circuit cases.
This sensor effectively rebuilds the unmeasured concentrations of
mRNA and the corresponding enzyme. Thus, the limitation of those experiments
in which only the concentration of the catalytically synthesized metabolite is available, can be overcome
by employing the simple software sensor applied here. This is a quite natural case if one takes into account that metabolites are quite stable at the molecular level.
At the same time, we can reproduce the concentrations of the unstable molecules of mRNA. This is a difficult task in experiments, despite the fact that the
mRNA dynamics has been partially or even totally unspecified.

The same scheme philosophy to build the observer is applied to a three-gene circuit with the purpose to show that the software sensor concept
could be in usage in a forward engineering approach. In this research however, we mentioned that we were able to show that the observer scheme designed in \cite{R. Aguilar y col. 2003}
for the single output case
works well also in a multiple variable case as embodied by a particular genetic circuit given in Fig.~(\ref{figureXX}).
The most stringent mathematical requirement for this extended applicability to the multiple output case is described below. The linear part of the dynamic system should be a matrix by blocks in which each of the blocks
should be of Brunovsky equivalent form. In addition, each subsystem corresponding to a superior block depends only on the subsystem corresponding to the next nearest block. This is a feature similar to
the property of Markoff processes. The Brunovsky equivalent form of the matrix blocks $A_i$ together with the structure of the corresponding output vector $C_i$ generate an observable pair $(A_i,C_i)$, giving us
the capability to infer the internal states of the gene network through the knowledge of its external outputs.
However, the special Brunovsky equivalent form of the blocks leads to the possible biological interpretation that each block of the linear part of the differential system represents only that contribution of the gene regulation mechanism that comes from reactions occurring in a cascade fashion. 

Another important issue that we tackled in this work is related to the way of adding the noise to the output of the dynamic system. Even though this is a typical situation from the standpoint of control process theory,
to the best of our knowledge it has not yet been applied in the biological context of gene regulation processes. We stress that this way of including noise effects could have both intrinsic and extrinsic
interpretations and therefore assure a more general approach of the noise problems. For example, in phenomenological terms, perturbations on the cells due to the measuring devices
and the experimental conditions,
together with the noise produced by the nature of the electronic instrumentation, could be equally described in this way.

In addition, this type
of nonlinear observer could be used as an online filter being
robust with respect to model uncertainties, i.e., neither
a known regulation function nor the parameter $K_{1,3}$ is required.

\bigskip

This work was sponsored in part by grants from the Mexican Agency {\em Consejo Nacional Ciencia y Tecnolog\'{\i}a} through  project 46980-R.


\end{document}